\documentclass[pdflatex,sn-nature]{sn-jnl}

\usepackage{graphicx}%
\usepackage{multirow}%
\usepackage{amsmath,amssymb,amsfonts}%
\usepackage{amsthm}%
\usepackage{mathrsfs}%
\usepackage[title]{appendix}%
\usepackage{xcolor}%
\usepackage{textcomp}%
\usepackage{manyfoot}%
\usepackage{booktabs}%
\usepackage{algorithm}%
\usepackage{algorithmicx}%
\usepackage{algpseudocode}%
\usepackage{listings}%
\usepackage{upgreek}
\usepackage{soul}

\theoremstyle{thmstyleone}%
\theoremstyle{thmstyletwo}%

\theoremstyle{thmstylethree}%

\begin{document}

\title[Article Title]{Doubly resonant nonlinear metasurfaces enabling NIR-to-UV upconversion for reconfigurable Fourier optical processing}


\author[1,2]{\fnm{Jumin} \sur{Qiu}}

\author[2,3]{\fnm{Meibao} \sur{Qin}}

\author*[1,4]{\fnm{Tingting} \sur{Liu}}\email{ttliu@ncu.edu.cn}

\author*[5]{\fnm{Lun} \sur{Qu}}\email{qulun@nankai.edu.cn}

\author[1]{\fnm{Xintong} \sur{Shi}}

\author[6]{\fnm{Feng} \sur{Wu}}

\author[2]{\fnm{Tianbao} \sur{Yu}}

\author*[1]{\fnm{Qiegen} \sur{Liu}}\email{liuqiegen@ncu.edu.cn}

\author*[1,4]{\fnm{Shuyuan} \sur{Xiao}}\email{syxiao@ncu.edu.cn}

\affil[1]{\orgdiv{School of Information Engineering}, \orgname{Nanchang University}, \orgaddress{\city{Nanchang}, \postcode{330031}, \country{China}}}

\affil[2]{\orgdiv{School of Physics and Materials Science}, \orgname{Nanchang University}, \orgaddress{\city{Nanchang}, \postcode{330031}, \country{China}}}

\affil[3]{\orgdiv{School of Education}, \orgname{Nanchang Institute of Science and Technology}, \orgaddress{\city{Nanchang}, \postcode{330108}, \country{China}}}

\affil[4]{\orgdiv{Institute for Advanced Study}, \orgname{Nanchang University}, \orgaddress{\city{Nanchang}, \postcode{330031}, \country{China}}}

\affil[5]{\orgdiv{The Key Laboratory of Weak-Light Nonlinear Photonics, Ministry of Education, School of Physics and TEDA Institute of Applied Physics}, \orgname{Nankai University}, \orgaddress{\city{Tianjin}, \postcode{300071}, \country{China}}}

\affil[6]{\orgdiv{School of Optoelectronic Engineering}, \orgname{Guangdong Polytechnic Normal University}, \orgaddress{\city{Guangzhou}, \postcode{510665}, \country{China}}}


\abstract{Fourier optical processing underpins optical information manipulation, yet extending such operations to short wavelengths within compact platforms remains challenging. Here, we address this challenge by embedding reconfigurable Fourier-domain processing within NIR-to-UV upconversion in a doubly resonant nonlinear metasurface. When coherently illuminated at the Fourier plane with an image-bearing signal and a spatially structured pump, the metasurface generates UV images via degenerate four-wave mixing. Crucially, the spatial-frequency content of these upconverted images is selectively shaped by the tailored spectrum of the pump. To boost the efficiency of this nonlinear process, the metasurface is designed to simultaneously support a toroidal dipole bound state in the continuum and a magnetic dipole resonance, providing spectrally aligned and independently enhanced field localization for signal and pump beams, respectively. Building on this architecture, we experimentally demonstrate directional and continuously tunable {filtering at the upconverted UV wavelengths}. These results establish nonlinear metasurfaces as a versatile platform for Fourier optics and reconfigurable all-optical image processing.}

\maketitle

\section{Introduction}\label{sec1}

Fourier optics provides a fundamental framework for manipulating spatial information encoded in optical fields and underpins a wide range of applications in image processing, pattern recognition, and optical information analysis\cite{goodman2017introduction,wetzstein2020inference}. Conventional Fourier-domain operations are typically implemented using free-space optical systems composed of lenses and spatial filters, where spatial-frequency components are accessed and selectively modified through linear propagation\cite{hu2024diffractive}. Although highly versatile, such systems are inherently bulky and difficult to integrate. Extending Fourier optical processing to short wavelengths, particularly in the UV regime, introduces further hurdles due to strong material absorption, a scarcity of high-quality optical components, and stringent alignment tolerances. These limitations underscore a pressing need for a compact platform capable of performing Fourier-domain manipulation at short wavelengths, beyond the capabilities of conventional linear architectures.

Metasurfaces, composed of planar arrays of subwavelength nanostructures, have emerged as a powerful platform to directly address the challenge of compactness, enabling flat-profile analogues of lenses, waveplates, and even spatial-frequency filters within deeply subwavelength thicknesses\cite{ren2017reconfigurable,zhou2020flat, zheng2021metasurface,wang2023metasurface,cotrufo2023polarization,deng2024poincare,liu2024edge,yu2026double}. However, directly operating at short wavelengths would require nanoscale feature sizes that push fabrication to its limits. Nonlinear metasurfaces offer an elegant bypass by intrinsically integrating frequency conversion with spatial light manipulation within a single ultrathin device, allowing to work with larger, easily fabricated structures at NIR wavelengths while achieving functionality in the UV regime through upconversion\cite{li2017nonlinear,Vabishchevich:23,wang2025all}. The inherently weak light–matter interaction at the subwavelength scale, however, necessitates resonant enhancement. High-quality-factor (Q) resonances, such as guided modes, Mie modes, and bound states in the continuum (BICs)\cite{hsu2016bound,huang2023resonant,PhysRevResearch.6.033073}, prove pivotal in this regard, strongly localizing optical fields and dramatically boosting the efficiency of nonlinear processes like third-harmonic generation (THG)\cite{liu2019high,koshelev2019nonlinear,xu2019dynamic,shi2020progressive,abdelraouf2024modal,liu2025polarization,sun2025high} and four-wave mixing (FWM)\cite{grinblat2017degenerate,liu2018all,colom2019enhanced,xu2022enhanced,moretti2024si,malek2025giant,franceschini2026intrapulse}.

Building on these advances, recent years have witnessed significant progress in nonlinear metasurface-based wavefront control and imaging\cite{grinblat2021nonlinear,zheng2023advances}. {Gradient metasurfaces employing geometric or propagation phases are widely used for nonlinear beam shaping and holography at upconverted wavelengths\cite{gao2018nonlinear,10.1021/acs.nanolett.0c01105,hail2024third,sedeh2025nonlinear,hu2026spin,cotrufo2026nonlinear,yao2026intensity,tian2026wavefront}.} The same design principle enables nonlinear metalenses that extend the reach of flat optics into the UV and even vacuum UV regimes\cite{tseng2022vacuum,zhou2022nonlinear,talts2025scalable,liu2025ultraviolet}. Concurrently, uniform metasurfaces are employed for pixel-level upconversion imaging through various resonant enhancement mechanisms, with schemes ranging from single-beam THG for narrowband to multiple-beam FWM for broadband image reconstruction\cite{zheng2023third,zheng2024broadband,sanderson2024infrared,liu2025high}. Despite these impressive demonstrations, the functionalities realized to date have mostly been confined to phase engineering or intensity mapping. {The broader realm of Fourier-domain optical processing on nonlinear metasurface platforms, encompassing operations such as spatial filtering, edge detection, and image differentiation, remains relatively underexplored.}

{Here, we take a step toward addressing this gap by demonstrating reconfigurable Fourier-domain image processing within doubly resonant enhanced degenerate four-wave mixing (dFWM) in an all-dielectric metasurface.} An image-bearing signal and a spatially structured pump are simultaneously incident onto the metasurface positioned in the Fourier plane, where the pump's tailored spectrum selectively shapes the Fourier components of the signal image. The dFWM directly implements a programmable spatial-frequency filter, accomplishing Fourier optical processing during frequency conversion. Critically, the metasurface supports two distinct high-Q resonances that are spectrally aligned with the signal and pump wavelengths, respectively, to independently enhance both incident fields, ensuring efficient dFWM to upconvert NIR inputs to UV outputs. By performing such Fourier-domain manipulation during nonlinear light generation, our approach unifies spatial-frequency processing and wavelength upconversion within a single ultrathin device, eliminating bulky linear filtering architectures and UV optical components. This work establishes nonlinear metasurfaces as a compact platform for Fourier optics, opening new avenues for all-optical image processing and short-wavelength photonic information manipulation.

\section{Results}\label{sec2}

\subsection{Overview}
\begin{figure*}[!ht]%
\centering
\includegraphics[width=\textwidth]{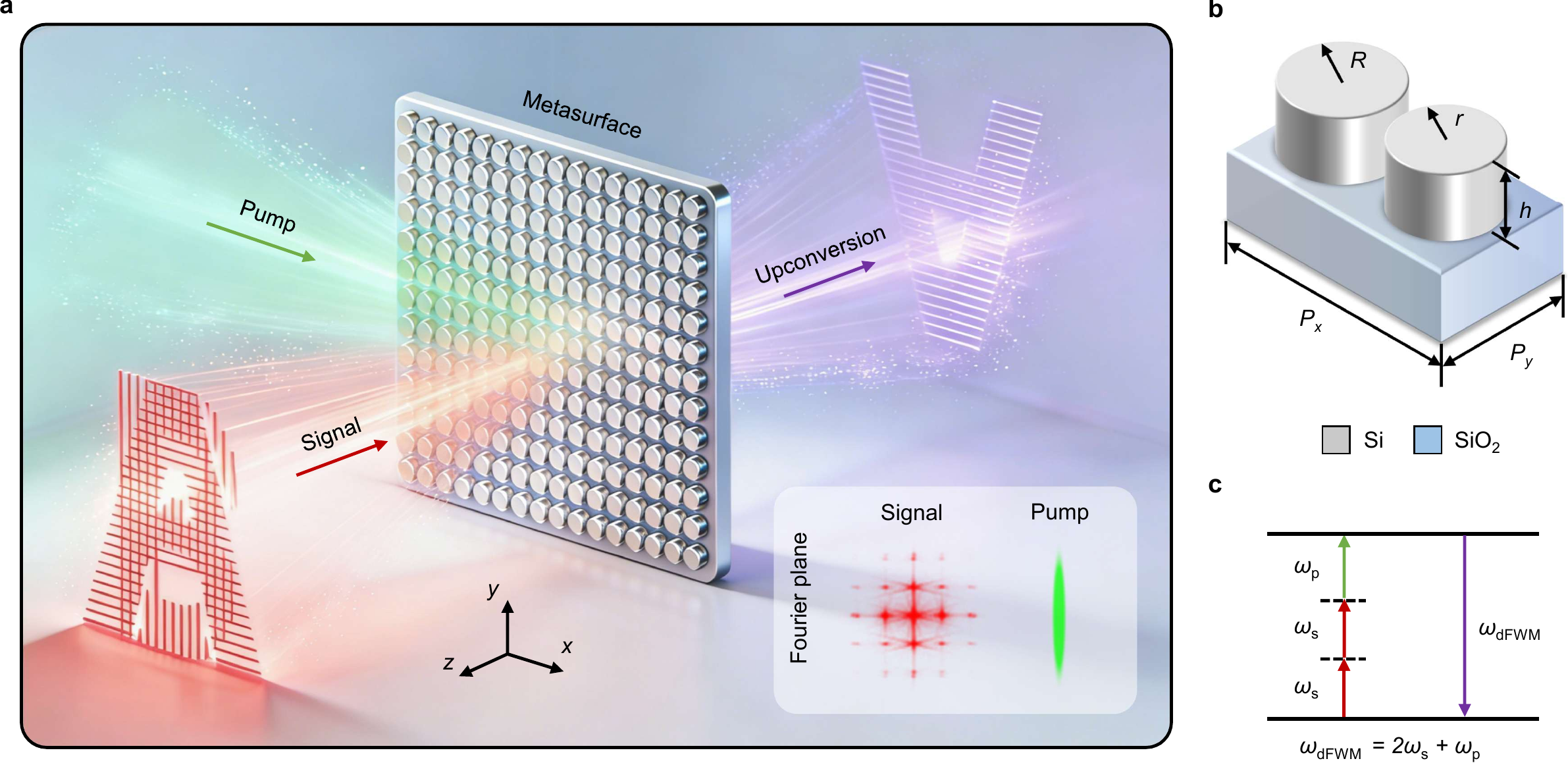}
\caption{\textbf{Principle of reconfigurable Fourier optical processing via a doubly resonant nonlinear metasurface.} \textbf{a} Conceptual illustration of the nonlinear Fourier processing platform. An image-bearing signal beam and a spatially structured pump beam are simultaneously focused onto the metasurface at the Fourier plane. The tailored pump spectrum selectively shapes the spatial-frequency components of the signal, generating a filtered UV image via dFWM. \textbf{b} Geometric design of the metasurface unit cell, featuring paired Si nanodisks (height $h = 220$ nm) arranged on a SiO$_2$ substrate with periods $P_{x} = 650$ nm and $P_{y} = 380$ nm. One nanodisk has a fixed radius $R = 135$ nm, while the other has a tunable radius $r$. \textbf{c} Energy-level diagram of the dFWM process $\omega_{\text{dFWM}} = 2\omega_{\text{s}}+\omega_{\text{p}}$ coupling the signal (s) and pump (p) fields. }\label{fig1}
\end{figure*}

{The working principle of our nonlinear Fourier processing platform is schematically illustrated in Fig. \ref{fig1}a.} A signal beam carrying spatial image information and a spatially structured pump beam are simultaneously focused onto a dielectric metasurface. The focusing lens performs an optical Fourier transform, mapping the object plane onto the metasurface; consequently, the field incident on the metasurface represents the spatial-frequency spectrum of the input image. In this Fourier plane, the structured pump beam exhibits a tailored spatial distribution, for instance, a slit-like profile that selectively transmits frequencies along a specific direction. Through the nonlinear optical process within the metasurface, an upconverted UV field is generated, whose frequency content is directly shaped by the pump's spatial spectrum before being transformed back to the image plane. As a conceptual demonstration, an input image consisting of a letter ``A'' formed by horizontal stripes and a letter ``B'' formed by vertical stripes can be filtered to retain only the horizontal components, yielding a distinct letter ``A'' in the upconverted UV output, illustrating how spatial-frequency filtering and wavelength conversion are accomplished in a single nonlinear step.

The nonlinear optical process underlying this operation is schematically depicted in Fig. \ref{fig1}c. The dFWM process $\omega_{\text{dFWM}} = 2\omega_{\text{s}}+\omega_{\text{p}}$, where subscripts $\text{s}$ and $\text{p}$ denote signal and pump, respectively, couples the two incident fields within the metasurface, generating a new optical field at the upconverted UV frequency. To enable this process efficiently, the metasurface unit cell comprises paired silicon (Si) nanodisks with slightly different radii, arranged periodically on a silica (SiO$_2$) substrate, as shown in Fig. \ref{fig1}b. Its structural asymmetry gives rise to two distinct high-Q resonances, which are spectrally aligned with the signal and pump wavelengths respectively. The resulting doubly resonant field confinement dramatically boosts the efficiency of the dFWM process, enabling the NIR-to-UV upconversion demonstrated in this work.

\subsection{Resonant mode analysis}

\begin{figure*}[!ht]%
\centering
\includegraphics[width=\textwidth]{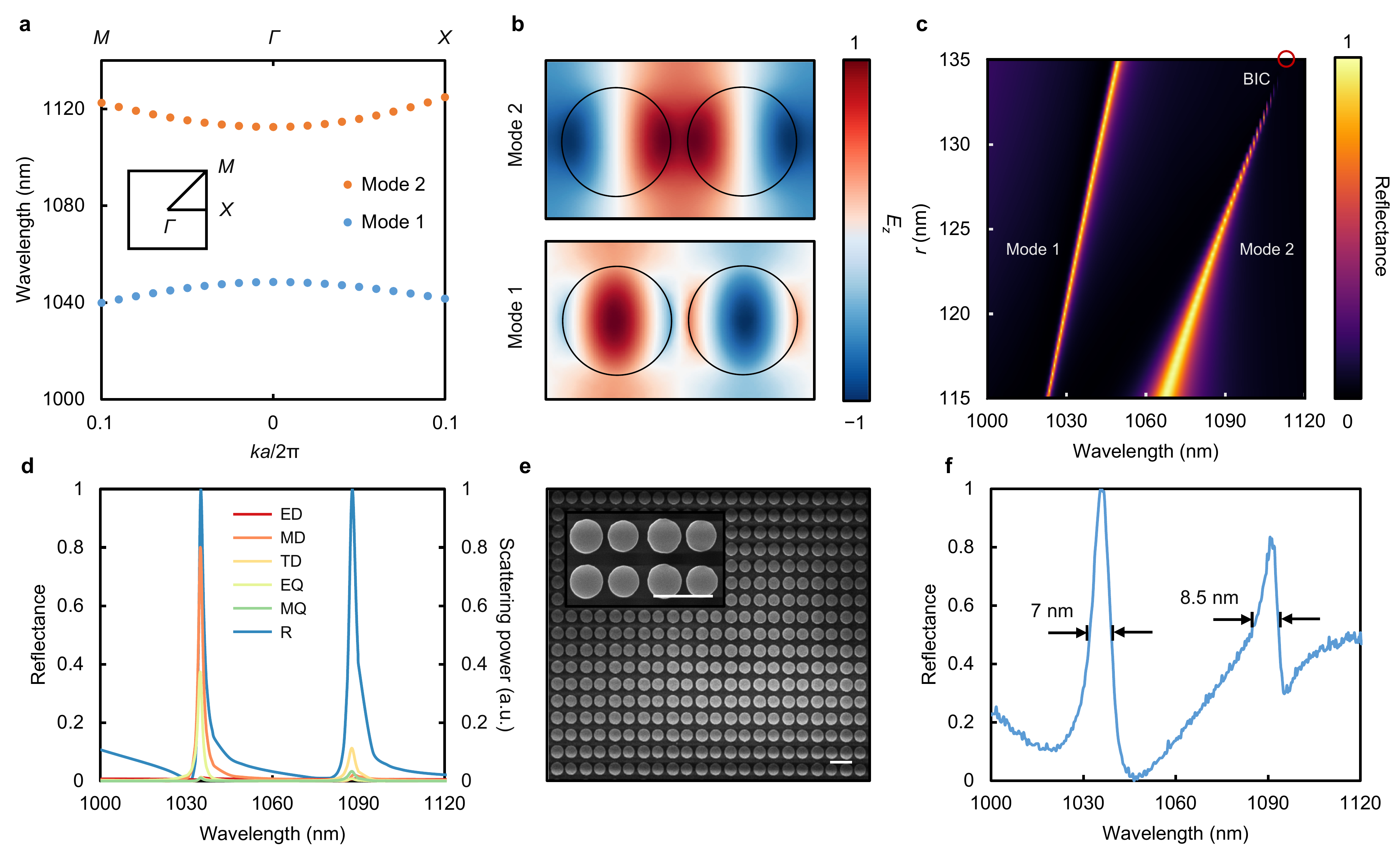}
\caption{\textbf{Resonant mode analysis and linear characterization of the metasurface.} 
\textbf{a} Calculated photonic band structure for a symmetric configuration ($r = R = 135$ nm).
\textbf{b} Simulated out-of-plane electric field ($E_{z}$) distributions of Mode 1 and Mode 2 at the $\Gamma$ point, corresponding to the bands in \textbf{a}.
\textbf{c} Calculated reflectance spectra under $x$-polarized incidence as a function of $r$.
\textbf{d} Multipole decomposition of the calculated reflectance for the asymmetric configuration ($r = 125$ nm).
\textbf{e} The SEM image of the sample. The scales are 500 nm.
\textbf{f} Experimentally measured reflectance spectrum of the fabricated sample ($r = 125$ nm). 
}\label{fig2}
\end{figure*}

To validate the design, we begin with a theoretical analysis to elucidate these resonant modes in the metasurface. {For the symmetric configuration, where the two nanodisks have identical radii ($r = R = 135$ nm), the photonic band structure along the high-symmetry $M$-$\Gamma$-$X$ direction in the first Brillouin zone is calculated, as shown in Fig. \ref{fig2}a. Two transverse-magnetic bands are observed in the wavelength range of interest, located near 1040 nm and 1120 nm. The corresponding eigenmodes are denoted as Mode 1 and Mode 2, respectively. At the $\Gamma$ point, Mode 1 exhibits a complex eigenfrequency with a finite imaginary component $f_1 = 285.92+0.33623i$ THz, indicating radiation leakage and thus a finite Q-factor. In contrast, Mode 2 possesses a purely real eigenfrequency $f_2 =$ 269.47 THz, signifying vanishing radiation loss and an ideally infinite Q-factor, a hallmark of BICs\cite{koshelev2018asymmetric,bogdanov2019bound,tu2026quasibound}. To further clarify the nature of these modes, we examine their out-of-plane electric field distributions $E_{z}$ at the $\Gamma$ point, as depicted in Fig. \ref{fig2}b. Under the $C_{2}$ rotational symmetry of the structure, Mode 1 exhibits odd parity and can couple to normally incident plane waves. Mode 2, however, displays even parity that is symmetry-incompatible with the incident field, suppressing radiation into the far field. This symmetry mismatch confirms that Mode 2 corresponds to a symmetry-protected BIC.}

To simultaneously excite both modes, we introduce structural asymmetry by varying the radius of one nanodisk from $r = 115$ nm to 135 nm while keeping the other fixed at $R = 135$ nm. The $x$-polarized plane wave is normally incident onto the metasurface, and the calculated reflectance spectra are shown in Fig. \ref{fig2}c. {As $r$ approaches $R$, the linewidth of Mode 2 progressively narrows, eventually vanishing in the symmetric limit, which is consistent with the formation of an ideal BIC\cite{https://doi.org/10.1002/adpr.202100144}.} In contrast, Mode 1 maintains a finite linewidth throughout the radius variation, in agreement with the eigenmode analysis results in Fig. \ref{fig2}a. {For a representative asymmetric configuration with $r = 125$ nm, we further decompose the reflectance into multipolar contributions, as presented in Fig. \ref{fig2}d. The resonance near 1030 nm (Mode 1) is dominated by the magnetic dipole (MD) component, while the resonance near 1090 nm (Mode 2) is predominantly governed by the toroidal dipole (TD) contribution, confirming the TD-dominated nature of the quasi-BIC. Other multipoles, including electric dipole (ED), electric quadrupole (EQ), and magnetic quadrupole (MQ), provide comparatively minor contributions. The coexistence of MD and TD-BIC resonances establishes the doubly resonant nature of the metasurface. For details on multipole decomposition and scattering power calculations, see Supplementary Note 1.}

The scanning electron microscopy (SEM) image of the fabricated sample and corresponding experimentally measured reflectance spectrum are shown in Figs. \ref{fig2}e and \ref{fig2}f. Two pronounced resonance peaks are observed at approximately 1030 nm and 1090 nm, in excellent agreement with the numerical simulations. {The measured angular dispersion follows the simulated trend, with Mode 1 shifting to shorter wavelengths and Mode 2 to longer wavelengths as the incidence angle moves away from normal (Supplementary Fig. S1). The linear reflection spectrum in Fig. 2f gives full widths at half maximum (FWHMs) of about 7 nm and 8.5 nm for the two resonances. By fitting the spectrum with temporal coupled mode theory\cite{fan2003temporal,wang2020controlling}, we extract Q-factors of approximately 147.7 and 128.4 for the two resonances, respectively.} These high-Q values confirm the strongly localized optical fields that underpins the nonlinear interactions demonstrated in the Nonlinear characterization section. The slight spectral deviations can be attributed to fabrication tolerances and material dispersion.

\subsection{Nonlinear characterization}

To exploit these doubly resonant enhancements for frequency conversion in the metasurface, we implement a dual-beam scheme. The pump beam is derived directly from a femtosecond fiber laser and spectrally fixed at 1034 nm to coincide with the MD resonance, while the signal beam is generated by an optical parametric amplifier and tuned to 1094 nm to align with the TD-BIC resonance. Under this doubly resonant excitation, we systematically characterize the generated nonlinear emissions, which include THG from each beam individually and dFWM from their interaction. 

\begin{figure*}[!ht]%
\centering
\includegraphics[width=\textwidth]{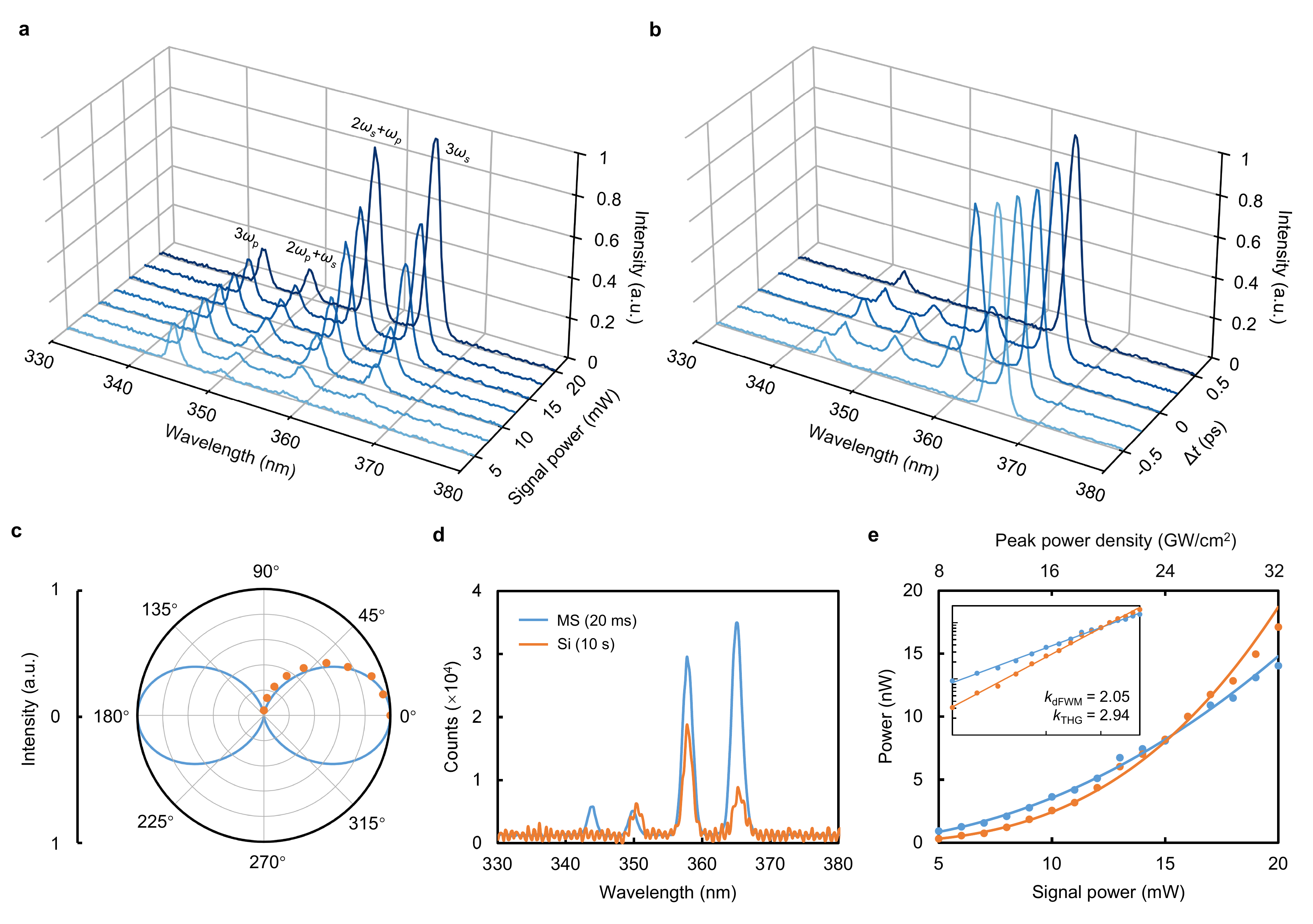}
\caption{\textbf{Nonlinear optical characterization of the metasurface.} 
\textbf{a} {Evolution of the nonlinear emission spectra as a function of the incident signal power at a constant pump power. The four UV channels, in order of increasing wavelength, are pump THG ($3\omega_p$), $2\omega_p+\omega_s$ mixing, the selected dFWM channel ($2\omega_s+\omega_p$), and signal THG ($3\omega_s$).}
\textbf{b} Nonlinear emission intensities versus the temporal delay between the pump and signal pulses.
\textbf{c} Polarization dependence of the selected dFWM emission intensity.
\textbf{d} Comparison of nonlinear emission spectra from the metasurface and an unpatterned bare Si film under identical doubly resonant excitation, {where the FWHM of each peak in the metasurface is approximately 2 nm.}
\textbf{e} {Measured dFWM (blue) and signal THG (orange) output powers as a function of incident signal power at a fixed pump power. The upper horizontal axis indicates the corresponding incident peak power density of the signal beam. Inset: logarithmic scale of the same data, with fitted slopes of 2.05 for dFWM and 2.94 for signal THG.}
}\label{fig3}
\end{figure*}

Figure \ref{fig3}a presents the evolution of the nonlinear spectrum as the signal power is gradually increased while keeping the pump power constant. When only the pump beam is present, a distinct THG peak at 344 nm ($3\omega_{\text{p}}$) is observed. Upon introducing the signal beam and increasing its power, additional nonlinear peaks emerge at 350 nm and 358 nm, corresponding to dFWM processes ($2\omega_{\text{p}}+\omega_{\text{s}}$ and $2\omega_{\text{s}}+\omega_{\text{p}}$) involving both pump and signal photons. Meanwhile, another THG peak at 365 nm ($3\omega_{\text{s}}$) associated with the signal beam also appears. At moderate signal power, the $2\omega_{\text{s}}+\omega_{\text{p}}$ component dominates the nonlinear spectrum. As the signal power is further increased, the $3\omega_{\text{s}}$ emission eventually surpasses the $2\omega_{\text{s}}+\omega_{\text{p}}$ output, reflecting the different power-scaling behaviors of these competing third-order nonlinear optical processes. 

To confirm the coherent nature of the dFWM processes, we measure the nonlinear emissions as a function of the temporal delay between the pump and signal pulses, as shown in Fig. \ref{fig3}b. Pronounced dFWM peaks at $2\omega_{\text{p}}+\omega_{\text{s}}$ and $2\omega_{\text{s}}+\omega_{\text{p}}$ are observed only when the delay approaches zero, where the two pulses overlap in time. When the delay exceeds approximately $\pm$ 0.5 ps, the dFWM emissions nearly vanish, confirming that the interaction requires the simultaneous presence of both pulses.

We further investigate the polarization dependence of the dFWM emission at $2\omega_{\text{s}}+\omega_{\text{p}}$. The pump polarization is fixed along the $x$-direction, while the signal polarization is rotated from 0° ($x$-polarization) to 90° ($y$-polarization). The intensity of the nonlinear peak at $2\omega_{\text{s}}+\omega_{\text{p}}$ decreases monotonically as the signal polarization deviates from 0°, vanishing almost completely at 90°. This trend reflects the polarization selectivity of the metasurface: the asymmetric configuration breaks the rotational symmetry, making the resonant modes predominantly excitable by $x$-polarized light. As the signal polarization rotates away from $x$, its projection onto the excitation axis diminishes, leading to a monotonic decrease in resonant enhancement and, consequently, a reduction in the dFWM emission intensity. The measured polar plot in Fig. \ref{fig3}c follows this trend closely. 

To quantify the resonantly enhanced nonlinear optical processes, we compare the generated nonlinear emissions from the metasurface with those from a bare Si film under identical excitation conditions, as shown in Fig. \ref{fig3}d. With both pump and signal powers set to 20 mW, the emission from the metasurface exhibits an approximately 800-fold enhancement for the dFWM process $2\omega_{\text{s}}+\omega_{\text{p}}$ and a nearly 1900-fold enhancement for the THG $3\omega_{\text{s}}$ compared to that from the unpatterned Si film. This dramatic improvement arises from the simultaneous field confinement at both interacting wavelengths, enabled by the MD and TD-BIC resonances. {Since the resonance linewidths of the metasurface are comparable to the laser spectral width of about 7 -- 9 nm, so most of the laser energy can be coupled into the designed resonant modes \cite{10.1021/acsphotonics.7b01478}. This efficient coupling contributes to the enhanced nonlinear optical processes.} Furthermore, the power dependence of these two nonlinear emissions is summarized in Fig. \ref{fig3}e. {The measured dFWM ($2\omega_{\text{s}}+\omega_{\text{p}}$) output power reaches 14.04 nW at incident signal and pump powers of 20 mW each, while the signal THG ($3\omega_{\text{s}}$) output power is 17.10 nW. The absolute conversion efficiency can be obtained from the formula $\eta_\text{dFWM}=P_\text{dFWM}/P_s$ and $\eta_\text{THG}=P_\text{THG}/P_s$ \cite{xu2022enhanced}, yielding values of $7.02 \times 10^{-7}$ and $8.55 \times 10^{-7}$, respectively. }The power of the dFWM emission scales quadratically with the incident signal power, exhibiting a slope of approximately 2 on a logarithmic scale, whereas the THG emission power scales cubically with a slope of approximately 3, which is consistent with the theoretical scaling law for the third-order nonlinear optical process\cite{liu2023high,qin2025enhanced}. {The slight deviation from power-law scaling at the highest signal powers may stem from local laser-induced damage that weakens the resonant field enhancement. A post-exposure optical micrograph shows local damage in the illuminated region (Supplementary Fig. S2).} The different power exponents explain the spectral evolution observed in Fig. \ref{fig3}a: at lower signal powers, dFWM dominates due to the strong pump contribution, while at higher signal powers, the cubic growth of THG eventually overtakes the dFWM output. Notably, all observed THG and dFWM emissions lie below 380 nm, confirming successful upconversion of the NIR light to the UV regime. This establishes the essential physical foundation for the cross-wavelength information processing at short wavelengths that will be demonstrated in the Fourier image filtering section.

\subsection{Fourier image filtering}

\begin{figure*}[!ht]%
\centering
\includegraphics[width=\textwidth]{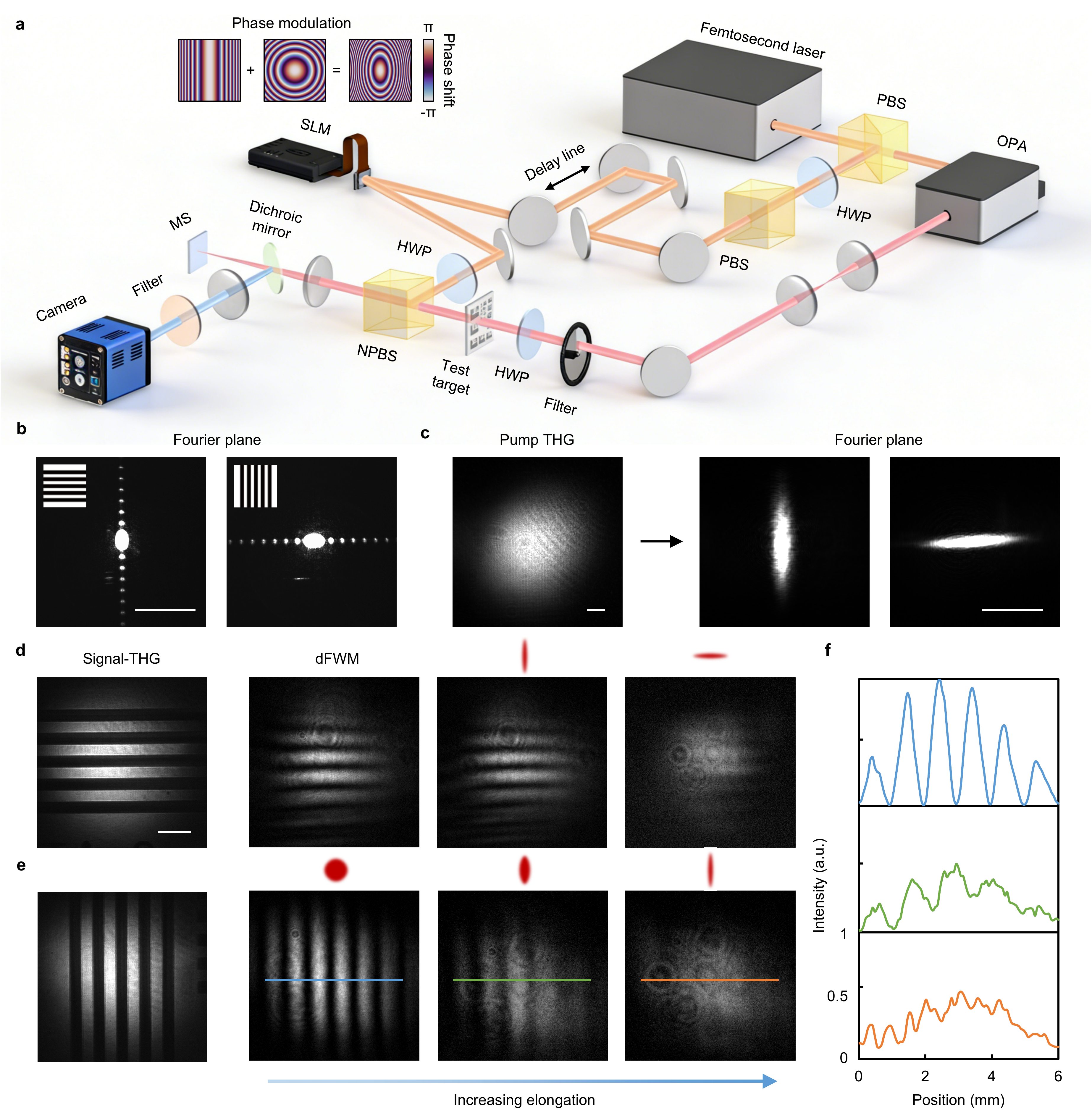}
\caption{\textbf{Reconfigurable Fourier-domain image filtering at upconverted UV wavelengths.} 
\textbf{a} Experimental setup for nonlinear Fourier image processing. The signal beam encodes an object image, and the pump beam is structured by an SLM. {Also shown are examples of programmable phase patterns applied to the SLM.}
\textbf{b,c} Measured intensity distributions of signal beam carrying the object image and the modulated pump beam in the Fourier plane.
\textbf{d} Directional spatial-frequency filtering of a horizontal stripe pattern. From left to right: signal THG reference image, and dFWM images obtained with circular, vertically elongated, and horizontally elongated pump beams.
\textbf{e} {Demonstration of continuously tunable filtering of a vertical stripe pattern, with the pump evolving from approximately circular to progress sively elongated along the same vertical axis. }
\textbf{f} {Extracted horizontal intensity profiles corresponding, in order, to the three dFWM images in \textbf{e}. }
All scale bars are 1 mm.
}\label{fig4}
\end{figure*}

To demonstrate the reconfigurable Fourier-domain image processing enabled by our doubly resonant metasurface, we construct the optical system illustrated in Fig. \ref{fig4}a. The signal beam passes through a resolution test target (NBS 1963A), encoding spatial image information onto the optical field. The pump beam is directed onto a spatial light modulator (SLM), where its phase profile is digitally programmed. The two beams are subsequently recombined collinearly and focused onto the metasurface by a lens. Owing to the Fourier-transform property of the focusing optics, the spatial distributions of both beams at the metasurface plane correspond to the spatial-frequency spectra of their respective input fields. The nonlinear optical process thus takes place directly in the Fourier domain. The generated UV field is collected by the same objective and projected onto a camera after appropriate spectral filtering, enabling direct visualization of the processed image. 

To tailor the spatial-frequency response, the pump beam is phase-modulated on the SLM. A programmable phase pattern combining a lens phase with a cylindrical lens phase is imposed. The lens term ensures proper focusing onto the metasurface plane, while the cylindrical component introduces one-dimensional spatial compression. As a result, the pump field in the Fourier plane is reshaped into a highly elongated distribution along a single spatial-frequency axis. {The measured intensity distributions of the signal beam carrying the object image and the modulated pump beam in the Fourier plane are shown in Fig. \ref{fig4}b and \ref{fig4}c, respectively. While the signal spectrum consists of discrete diffraction spots, the pump is reshaped into a continuous, slit-like spatial-frequency profile. In addition, fine alignment was achieved by superimposing a tilt phase on the phase pattern displayed on the pump SLM, allowing its position to be finely adjusted relative to the signal in the Fourier plane.} Since the dFWM process depends on the product of the interacting fields, this elongated pump distribution effectively acts as a directional Fourier filter: spatial-frequency components aligned with the elongated axis are preferentially amplified in the generated UV field, while orthogonal components are suppressed. {The incident peak power densities at the metasurface were approximately 30 and 15 GW/cm$^2$ for the pump and signal beams, respectively.}

We first evaluate the filtering performance using a horizontal stripe pattern from the resolution target. The results are presented in Fig. \ref{fig4}d. For comparison, the THG image of the signal beam (generated via $3\omega_{\text{s}}$) is shown as a reference, representing direct wavelength upconversion without Fourier filtering. The dFWM image generated with a circular pump (via $2\omega_{\text{s}}+\omega_{\text{p}}$) preserves the global structure of the object, yet displays reduced sharpness and contrast relative to the THG reference. This behavior reflects the multiplicative nature of the FWM process in momentum space: spatial-frequency components with intrinsically lower amplitudes, particularly at higher frequencies, contribute less efficiently to the nonlinear polarization, leading to preferential attenuation of fine features in the reconstructed image. In this sense, the dFWM naturally acts as a soft low-pass filter. Building on this baseline response, we then apply anisotropic pump shaping to realize directional spatial-frequency filtering. {When the pump beam is vertically elongated in the Fourier plane, the vertical spatial-frequency components associated with the horizontal stripes are preferentially retained in the upconverted image. Conversely, a horizontally elongated pump reduces its overlap with these components, suppressing the horizontal stripe modulation.} These results clearly demonstrate that image filtering is achieved during the nonlinear optical process itself, rather than through post-detection processing.

To further illustrate the tunability of the filtering response, we perform experiments using a vertical stripe pattern, as shown in Fig. \ref{fig4}e. By gradually varying the focal length of the cylindrical phase term programmed onto the SLM, the pump distribution in the Fourier plane evolves continuously from circular to strongly elongated. Correspondingly, the filtering effect becomes progressively more pronounced: as the pump beam becomes increasingly elongated, orthogonal spatial-frequency components are progressively suppressed in the dFWM image. To quantify this evolution, we extract horizontal line profiles from the images in Fig. \ref{fig4}e and plot the corresponding intensity distributions in Fig. \ref{fig4}f. The contrast between transmitted and suppressed spatial features increases with pump elongation, providing quantitative evidence that the Fourier filtering strength can be continuously tuned via digital modulation of the pump phase.

{These results demonstrate the pump-controlled filtering principle illustrated in Fig. \ref{fig1}a. Changing the pump orientation selects which spatial-frequency components are retained, while varying the pump elongation at a fixed orientation tunes the stripe modulation in the upconverted UV image.} This establishes nonlinear metasurfaces as a reconfigurable platform for optical processing and short-wavelength image manipulation.

\section{Discussion}

We have demonstrated a doubly resonant nonlinear metasurface that integrates reconfigurable Fourier-domain image processing with wavelength upconversion on a compact platform. By independently enhancing the signal and pump fields via a TD-BIC and an MD resonance, we achieve efficient dFWM that upconverts NIR to UV. Because this nonlinear optical process occurs in the Fourier plane, the generated UV field inherently carries the result of spatial-frequency filtering imposed by the structured pump. This stands in contrast to conventional approaches, where the pump beam serves merely as a uniform control knob\cite{shcherbakov2015ultrafast,pogna2021ultrafast,di2024all,bijloo2026all}. Here, we transform the structured pump into an active participant in optical computing, with its tailored spectrum directly defining a programmable filter in the Fourier domain. The directional and continuously tunable filtering achieved in our experiments verifies the viability of this generation-stage processing paradigm. By unifying Fourier-domain manipulation and upconversion within a single ultrathin metasurface, we directly address the need for a compact platform capable of short-wavelength operation, eliminating bulky linear spatial filters and external UV optics. {The same principle can be extended to more sophisticated operations, such as edge detection or convolution, by programming appropriate pump phases\cite{zheng2022meta,luo2022metasurface,luo2024meta,Li:18}, and to other spectral regimes through suitable material and geometric scaling\cite{del2021infrared,kruk2022asymmetric,zhu2022nonlinear,valencia2024enhanced,zheng2024mid,ma2025quantum,qu2025beam,cotrufo2025nonlinear}. 
In addition, directly controlling the metasurface response via electrical, thermal, or optical means could offer a potential route to tuning Fourier processing\cite{Cotrufo2024,dias2026allopticalcontrolnonlocalityultrafast}, for example by enabling direct adjustment of the resonant response and thereby complementing the spatial control provided by the pump.}
Our results highlight nonlinear metasurfaces as a promising platform for Fourier optics and short-wavelength image processing.

\section{Methods}

\subsection{Numerical simulation}
Numerical simulations of the metasurface are performed using the finite-element method implemented in COMSOL Multiphysics. The refractive index of Si is taken from experimental data reported by Palik\cite{palik1998handbook}, while the SiO${_2}$ substrate is modeled with a constant refractive index of 1.46. The simulation domain consists of a single unit cell, which is treated as part of an infinite periodic array by applying Floquet periodic boundary conditions to the in-plane boundaries. The substrate is modeled as semi-infinite, and perfectly matched layers (PMLs) are employed along the propagation direction to absorb outgoing waves and minimize spurious reflections. For the eigenmode analysis, the eigenfrequency solver is used to calculate the photonic band structure, resonant eigenfrequencies, and the out-of-plane electric field distributions of the modes. For the reflectance spectrum calculations, ports are placed at the inner interfaces of the PMLs to excite normally incident plane waves and to extract reflectance from the scattering parameters. Multipole decomposition of the resonant modes is performed by integrating the induced current densities within the unit cell, allowing quantitative identification of electric, magnetic, and toroidal multipole contributions.

\subsection{Sample fabrication}
The metasurface samples are fabricated on a Si-on-insulator (SOI) wafer consisting of a 220 nm-thick device layer, a 2 $\upmu$m buried oxide layer, and a 700 $\upmu$m Si substrate. A positive-tone electron-beam resist (ZEP520) is spin-coated onto the cleaned SOI wafer, and electron-beam lithography (EBL) is employed to define the designed nanopatterns. After development, a 30 nm chromium (Cr) layer is deposited onto the patterned resist by electron-beam evaporation. A lift-off process in N-methyl-2-pyrrolidone (NMP) removes the residual resist along with the Cr on top of it, leaving behind Cr hard masks that replicate the original EBL patterns. The patterns are then transferred into the Si device layer by inductively coupled plasma reactive ion etching (ICP-RIE) using a SF$_6$/C$_4$F$_8$ gas mixture. After etching, the remaining Cr mask is removed by a commercial Cr etchant.  {The schematic of the fabrication process flow chart of the metasurfaces is shown in Supplementary Fig. S3,} and the SEM image of a representative structure with $R=135$ nm and $r=125$ nm is shown in Fig. \ref{fig2}e. 

\subsection{Experimental setup}
{For linear angle-resolved reflectance measurement, a halogen lamp (DC-950H, Dolan-Jenner) is used as the broadband light source. The output beam is first passed through a Köhler illumination system to control the illumination angle and illuminated area, and is then prepared in a specific polarization state using a polarizer, a half-wave plate (HWP), and a quarter-wave plate (QWP).  The beam then passes through a beam splitter (BS) and is focused by an objective lens (M Plan Apo NIR 20×, Mitutoyo) to illuminate the metasurface sample. The reflected light is collected by the same objective and directed by a second BS, which splits it into two paths: one portion is routed to an imaging spectrometer (Omni-$\lambda$300i, Zolix) via a series of relay lenses for spectral analysis, where Fourier-space detection enables rapid acquisition of continuous spectral information over a wide angular range. The remaining portion is captured by a camera (MV-CI013-GS-TH, Hikrobot) for real-time imaging and alignment of the sample surface. The schematic of the measurement setup is shown in Supplementary Fig. S4.}

For nonlinear optical characterization and imaging, a femtosecond fiber laser (FemtoYL-40, YSL Photonics) delivers pulses at a central wavelength of 1034 nm, with a pulse duration of 200 fs and a repetition rate of 200 kHz. The beam is initially split into two paths by a polarizing beam splitter (PBS). The reflected path serves as the pump beam, with its power adjusted by a combination of a HWP and a PBS. This path incorporates a delay line composed of multiple mirrors to control the pulse timing, followed by a spatial light modulator (PLUTO-2.1-NIR-145, Holoeye) for phase programming and a HWP for polarization control. The transmitted path, serving as the signal beam, is first tuned to the desired wavelength using an optical parametric amplifier (ParaAMP-8, Parametric Photonics). A 4f optical system is employed to adjust the beam divergence, followed by a variable neutral density filter for power control and a HWP for polarization adjustment. An NBS 1963A test target (RTS2AE-P, LBTEK) serves as the imaging object. The pump and signal beams are subsequently recombined by a non-polarizing beam splitter (NPBS) and focused onto the metasurface sample. The generated nonlinear emissions are separated from the residual excitation beams using a dichroic mirror. These emissions are then directed through a collection lens and imaged onto a camera (pco.edge 5.5, PCO), with specific spectral components isolated by bandpass filters. Alternatively, the nonlinear emission can be fiber-coupled to a UV-NIR spectrometer (USB4000, Ocean Optics) for spectral characterization. To ensure measurement accuracy, the spectrometer is radiometrically calibrated using a calibrated light source (HL-2000-CAL, Ocean Optics). {The fold enhancements in Fig. \ref{fig3}d are calculated from background-subtracted count rates normalized to the respective exposure times. The peak signal-to-noise ratio (SNR) is 51.3 dB for the metasurface and 35.5 dB for bare Si, indicating that noise is not the limiting factor in the enhancement estimate.}

For power measurements, two movable power detectors are employed. A germanium detector (918D-ST-IR, Newport) is placed after the objective to measure the NIR power incident on the metasurface, while a silicon detector (918D-ST-UV, Newport) is positioned after the dichroic mirror to measure the power of the generated nonlinear emission in the UV range. Both detectors are connected to a dual-channel optical power meter (2936-R, Newport) for real-time synchronous readout. The schematic of the power measurement setup is shown in Supplementary Fig. S5.

\section*{Data availability}

Data underlying the results are available from the corresponding authors upon request.



\section*{Acknowledgments}

This work was supported by the National Natural Science Foundation of China (Nos. 12304420, 12264028, 12364045, 12364049, 12547140, and U24A20304), the Natural Science Foundation of Jiangxi Province ((Nos. 20242BAB25041, 20262BAC240253, and No. 20253BAC260002), the Young Elite Scientists Sponsorship Program by JXAST (Nos. 2023QT11 and 2025QT04).

\section*{Author Contributions}

J.Q. and S.X. conceived the idea. M.Q., T.L., and X.S. performed the theoretical calculations and numerical simulations. J.Q. and S.X. conducted the experiments. J.Q. wrote the original draft. T.L., L.Q., F.W., and T.Y. reviewed and edited the manuscript. T.L., L.Q., Q.L., and S.X. supervised the project. All authors discussed the results and approved the final version of the manuscript.

\section*{Competing Interests}

The authors declare no competing interests.

\end{document}